\documentclass{osa-article}

\usepackage{upgreek}
\usepackage{float}
\usepackage{siunitx}
\usepackage{xcolor}
\usepackage{tikz, pgf}
\usepackage{setspace}
\usepackage[verbose]{placeins}
\usetikzlibrary{positioning, shapes, arrows}
\usepackage{hyperref}

\pgfdeclarelayer{background}
\pgfdeclarelayer{foreground}
\pgfsetlayers{background,main,foreground}

\journal{oe}

\articletype{Research Article}

\begin{document}

\title{Pockels-effect-based adiabatic frequency conversion in ultrahigh-\textit{Q} microresonators}

\author{Yannick Minet,\authormark{1,2,4} Luis Reis,\authormark{1,4} Jan Szabados,\authormark{1} Christoph S. Werner,\authormark{3} Hans Zappe,\authormark{2} Karsten Buse,\authormark{1,3} and Ingo Breunig\authormark{1,3,*}}

\address{\authormark{1}Laboratory for Optical Systems, Department of Microsystems Engineering - IMTEK, University of Freiburg, Georges-K{\"o}hler-Allee 102, 79110 Freiburg, Germany\\
\authormark{2}Gisela and Erwin Sick Chair of Micro-Optics, Department of Microsystems Engineering - IMTEK, University of Freiburg, Georges-K{\"o}hler-Allee 102, 79110 Freiburg, Germany\\
\authormark{3}Fraunhofer Institute for Physical Measurement Techniques IPM, Heidenhofstra{\ss}e 8, 79110 Freiburg, Germany\\
\authormark{4}These authors contributed equally to this work.}

\email{\authormark{*}optsys@ipm.fraunhofer.de}


\begin{abstract}
%\begin{doublespace}
Adiabatic frequency conversion has some key advantages over nonlinear frequency conversion.
No threshold and no phase-matching conditions need to be fulfilled.
Moreover, it exhibits a conversion efficiency of $100\,\%$ down to the single-photon level.
Adiabatic frequency conversion schemes in microresonators demonstrated so far suffer either from low quality factors of the employed resonators resulting in short photon lifetimes or small frequency shifts.
Here, we present an adiabatic frequency conversion (AFC) scheme by employing the Pockels effect.
We use a non-centrosymmetric ultrahigh-\textit{Q} microresonator made out of lithium niobate. Frequency shifts of more than $5\,$GHz are achieved by applying just $\SI{20}{\volt}$ to a $70{\text -}\si{\micro\meter}$-thick resonator.
Furthermore, we demonstrate that with the same setup positive and negative frequency chirps can be generated.
With this method, by controlling the voltage applied to the crystal, almost arbitrary frequency shifts can be realized.
The general advances in on-chip fabrication of lithium-niobate-based devices make it feasible to transfer the current apparatus onto a chip suitable for mass production.
%\end{doublespace}
\end{abstract}

%%%%%%%%%%%%%%%%%%%%%%%%%%  body  %%%%%%%%%%%%%%%%%%%%%%%%%%
%\begin{doublespace}
\section{Introduction}
Optical frequency conversion in microresonators has been advanced over the last decades.
The vast majority of practical realizations are based on the nonlinear response of the material to light\,\cite{Boyd.2008}.
For example, frequency combs in microresonators made out of centrosymmetric materials \cite{Herr14b} and tunable optical parametric oscillators in non-centrosymmetric microresonators have been demonstrated \cite{Furst10,Breunig16}.
High conversion efficiencies require high intensities, as well as  phase matching conditions need to be fulfilled\,\cite{Boyd.2008}.
Moreover, a pump threshold must be overcome for the most versatile conversion mechanism, optical parametric oscillation.
An alternative optical conversion technique is the so-called adiabatic frequency conversion (AFC).
Here, the frequency of light traveling in a resonator is shifted due to a change of the optical round-path length.
One implementation is to change the refractive index of the material and to keep the geometrical path length constant.
The frequency of light changes then according to \cite{Notomi06}
\begin{equation}
\Delta\nu \approx -\nu\frac{\Delta n}{n}.
\label{eq:refractivvsfrequency}
\end{equation}
The change of the refractive index must happen in a time $\Delta t$ shorter than the propagation time $t$ of the light in the material, i.e. before it gets lost by absorption or scattering.
Microresonators act as light traps that can store light for many nanoseconds or even milliseconds, depending on their quality factor $Q$. Thus, they are well suited to realize AFC\,\cite{Savchenkov07e}.

For AFC, no threshold has to be overcome. This frequency conversion scheme has been realized experimentally in several implementations, even down to a single photon level \cite{Preble.2012,Fan16b}.  
For example AFC was shown in photonic crystals\,\cite{Tanabe.,Tanabe.b,Kondo18}, in waveguides \cite{Kampfrath10,Upham.2010,Fan16b,Kondo.14} and in fiber grating cavities\,\cite{Kabakova12}.
So far, two different attempts involve microresonators:
Conduction-band electrons generated by laser pulses, inducing a change of the refractive index allow to shift a few hundred GHz to shorter wavelengths\cite{Preble.b}.
However, the quality factor suffers from the generated free electrons and hence losses increase.
Typical $Q$ values are around $10^4$, only.
Moreover the repetition rate is limited by the time it takes for the free electrons to recombine.
An alternative method is to shift the frequency adiabatically to longer wavelengths via the AC-Kerr effect \cite{Yoshiki.}.
Here, a second pump laser is needed.
Without affecting the quality factor being as high as $Q\approx 10^7$, frequency shifts of hundreds of MHz can be observed.
Thus these approaches come either with low $Q$-factors and large tuning ranges or high-$Q$ and relatively small tuning range.
\\ 
Here we demonstrate adiabatic frequency conversion by employing the linear electro-optic effect, the so-called Pockels effect, in high-$Q$ non-centrosymmetric microresonators.
In contrast to the carrier-induced method we can take advantage of a linear dependence of the tuning on the applied electric field.
The Pockels effect gives the opportunity to create both positive and negative frequency shifts, which is also not possible with approaches demonstrated so far.
This allows us to create arbitrary wavelength-vs.-time functions, since the electric signal is transferred directly into a wavelength change.

At first sight our approach is similar to up- and down conversion of microwave photons by employing the Pockels effect.
Here, high conversion efficiencies from the microwave to the optical domain were achieved\cite{Faneaar4994,Rueda:16}.
In contrast to AFC, to get high conversion efficiencies, considerably larger optical pump power is needed, as it is the case, e.g., for single photon-up-conversion\cite{dam2012room}.  
Another difference between electro-optic modulation (EOM) in microresonators and AFC is the fact that for EOM experiments the modulation frequency determines the frequency shift\cite{Rueda:16,rueda2019resonant,Zhang.2019}. For adiabatic frequency conversion the frequency shift depends on the applied voltage.
EOM in microresonators can be considered as a non-adiabatic transition: here, energy is transferred between neighboring resonator modes\cite{dong2008inducing,Daniel:11}. A suited spatial overlap of the optical mode and the electric field is in both cases necessary to achieve efficient conversion\cite{PhysRevA.94.053815}.

\section{Theoretical considerations}
\begin{figure}[htbp!]
\centering
\includegraphics{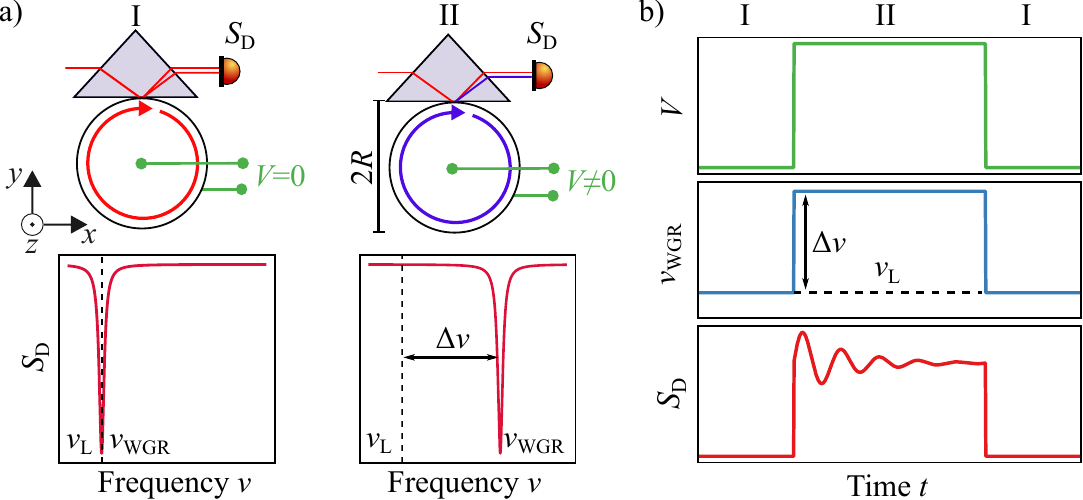}
	\caption{a) Photodiode signal $S_\textrm{D}$ showing the transmission spectrum of a WGR. In (I), the situation with no electric field applied and laser light coupled into a resonator mode is shown. In (II) an electric field is applied faster than the decay time $\tau$ between the electrodes on the +$z$- and $-z$-faces of the resonator: due to the Pockels effect-induced change $\Delta\nu$ of the eigenfrequency $\nu_\mathrm{WGR}$, no light is coupled into the resonator anymore.
	b) By applying a square-function electrical signal, the out-coupled light changes its frequency $\nu_{\textrm{WGR}}$, while the laser frequency $\nu_{\textrm{L}}$ stays constant.
	The signal $S_\textrm{D}$ on the photodiode, shows an exponentially decaying oscillation.
	The rise/drop of the base level $S_\textrm{D}$ when the voltage is switched on/off stems from the fact that with the voltage applied less pump light is coupled into the resonator and hence reflected.
	Thus the light power reaching the photodiode is increased.}
	\label{fig:tvsVFT}
\end{figure}

Optical whispering-gallery-resonators (WGRs) are monolithic  spheroidally-shaped devices that can provide long photon lifetimes and tight spatial confinement of light.
Light is coupled into the resonator when its frequency $\nu_\mathrm{L}$ matches the eigenfrequency $\nu_\mathrm{WGR}$ of the resonator, which can be calculated as
\begin{align}
    \nu_\mathrm{WGR} \approx m \frac{c}{2\pi R n}
    \label{eigenfrequency}
\end{align}
where $R$ is the major radius of the microresonator (Fig.~\ref{fig:tvsVFT}a)), $c$ is the vacuum speed of light and $n$ is the  refractive index of the bulk material.
The number $m$ represents the azimuthal mode number \cite{Breunig16}.
 Another important parameter is the quality factor $Q$ corresponding to the number of oscillations the electric field of light can travel around the WGR rim until it decays due to losses. The decay time $\tau$ until the light intensity in the WGR decreases to $e^{-1}$ of its initial value, is directly proportional to the quality factor \cite{Preble.b}:
\begin{equation}
    \tau = \frac{Q}{2\pi\nu_\mathrm{WGR}}.
    \label{decaytime}
\end{equation}
%The Q is also inversely proportional to the $Q=\nu/\Delta\nu_\mathrm{FWHM}$.}
In adiabatic frequency conversion, as illustrated in Fig.\ref{fig:tvsVFT}, the eigenfrequency $\nu_\mathrm{WGR}$ is changed during this time, and the frequency of the trapped light is forced to follow this change. In this contribution, we change the refractive index $n$ by the Pockels effect according to \cite{Saleh2007}
\begin{align}
	\Delta n = -\frac{1}{2}  n^3 r E.
	\label{eq:pockels}
\end{align}
where $r$ is the Pockels coeffcient and $E$ the externally applied electrical field.
To calculate the shift $\Delta\nu$ of the eigenfrequency $\nu_{\textrm{WGR}}$ as shown in Fig.~\ref{fig:tvsVFT} a), we insert Eq.~(\ref{eq:pockels}) in Eq.~(\ref{eq:refractivvsfrequency}) to obtain
\begin{equation}
\Delta \nu = \frac{1}{2} \nu_{\textrm{WGR}}n^2 r E_z,
\label{eq:frequency_shift}
\end{equation}
where $E_z$ is the electric field in $z$-direction.
If we apply a voltage along the $z$-direction of a $z$-cut congruent lithium niobate (CLN) crystal as seen in Fig.\ref{fig:tvsVFT}a), we address two $r$-coefficients (Pockels coefficients): for light polarized parallel to the $z$-direction (e-pol.) we access $r_{33}$ and for light polarized perpendicular to the $z$-direction (o-pol.), we access $r_{13}$. To estimate the expected eigenfrequency shift for an applied field of $E_{z}=10$~V/mm, we consider $r_{33}\approx\SI[per-mode=symbol]{24.6}{\pico\meter\per\volt}$ and $r_{13}\approx \SI[per-mode=symbol]{8.1}{\pico\meter\per\volt}$ for undoped CLN around \SI{1000}{\nano\meter}\,\cite{Mendez99b}.
If we consider a cavity which is resonant for light with a wavelength of $\lambda_\mathrm{L}= \SI{1040}{\nano\meter}$ we obtain a shift of $\Delta\nu_\mathrm{e}\approx\SI{165}{\mega\hertz}$ and $\Delta\nu_\mathrm{o}\approx\SI{58}{\mega\hertz}$.
Consequently, we expect to see three fold larger tuning for e-pol. than for o-pol. light.

The electric field also leads to a major radius change $\Delta R$ due to the piezo-electric effect. The corresponding expansion is \begin{equation} \Delta R = d_{311}RE_{z}, \end{equation} where $d_{311}=-1$~pC/N, according to \cite{Weis85}. As can be seen from Eq.~(\ref{eigenfrequency}), this also leads to eigenfrequency tuning, which can, 
analogously to Eq.~(\ref{eq:refractivvsfrequency}), be calculated as \begin{equation} \Delta\nu_\mathrm{WGR,piezo}\approx -\nu_\mathrm{WGR}\frac{\Delta R}{R}.
\end{equation}
Thus, for a typical resonator with $R=1$~mm, the application of $E_{z}=10$~V/mm would lead to an eigenfrequency shift of $\Delta\nu_\mathrm{WGR,piezo}\approx3$~MHz only for light with an initial wavelength of $\lambda_\mathrm{L}=1040$~nm regardless of its polarization. This is more than one order of magnitude lower than the eigenfrequency shift induced by the Pockels effect and is thus neglected in the following.

To detect AFC, we use the scheme shown in Fig.~\ref{fig:tvsVFT}.
On the photodiode we expect to observe a beat signal $S_{\textrm{D}}(t)$ between the out-coupled light and the original pump laser light.
The signal is supposed to be described by
\begin{align}
S_{\textrm{D}}(t) &=I_{\textrm{L}}+I_{\textrm{WGR}}\exp{\left( -\frac{t}{\tau}\right)} \notag\\
&+2 \sqrt{I_{\textrm{L}} I_{\textrm{WGR}}}\cos(2\pi\Delta\nu t + \phi_0)\exp{\left( -\frac{t}{2\tau}\right)}.
\label{eq:theosignal}
\end{align}
Here $I_{\textrm{L}}$ is the intensity of the pump laser that is reflected at the prism base, $I_{\textrm{WGR}}$ is the intensity of the light leaving the WGR via the coupling prism and $\phi_0$ is the relative phase between the out-coupled light and the laser light. 
The beat note has the frequency $\Delta\nu = |\nu_{\tiny\textrm{L}}-\nu_{\tiny\textrm{WGR}} |$, with $\nu_{\tiny\textrm{L}}$ being the frequency of the laser light originally coupled into the resonator and  $\nu_{\tiny\textrm{WGR}}$ the frequency of the manipulated light leaving the WGR.

\section{Methods}
\begin{figure}[htbp!]
	\centering
    \includegraphics{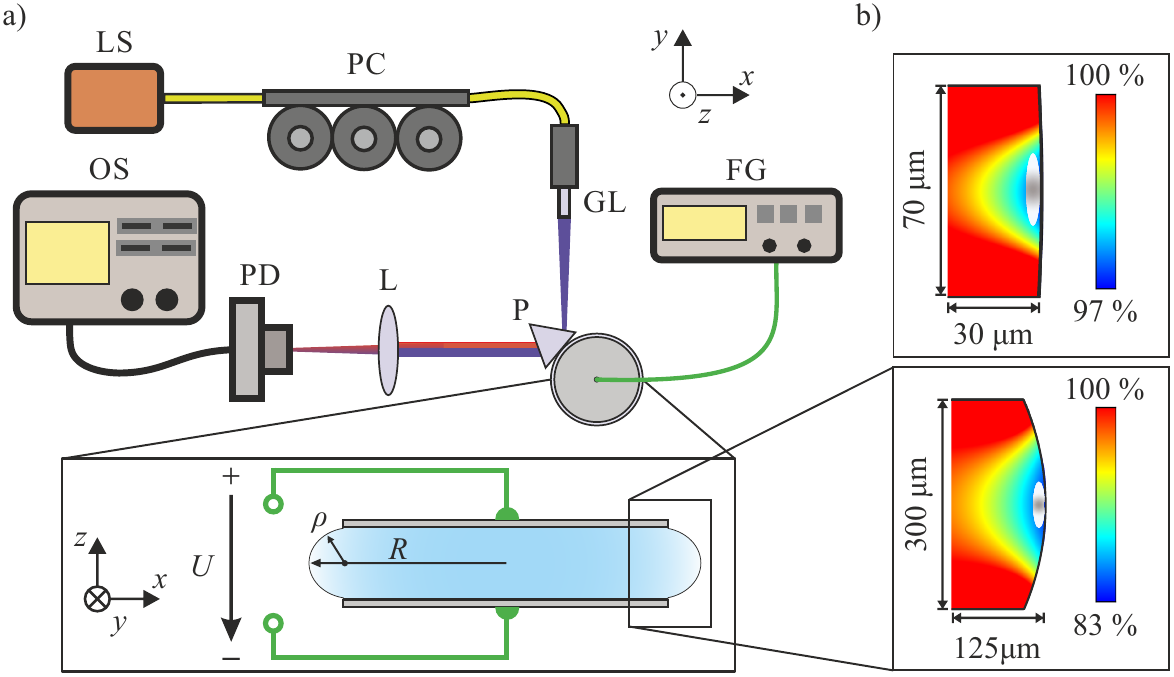}
	\caption{a) Schematic of the measurement setup. Abbreviations: Tunable laser (LS); polarization controller (PC); GRIN lens (GL); function generator (FG); lens (L); prism (P); photodiode (PD); oscilloscope (OS).
	b) The close up shows the finite element simulation of the electric field $E_z$ in the resonator using our real geometries.
	The small gray area represents, as a guide to the eye, the volume and position of the fundamental resonator mode.
	Here $R$ and $\rho$ are the major and the minor radius of the resonator, respectively.}
	\label{fig:setup}
\end{figure}

The resonators used in this work are fabricated out of a $300\text{-}\si{\micro\meter}$-thick wafer of $z$-cut 5-\%-MgO-doped congruent lithium niobate (CLN).
Both sides of the wafer are covered with a $\SI{150}{\nano\meter}$-layer of chromium.
First we cut out  a resonator blank using frequency-doubled $\SI{388}{\nano\meter}$ femtosecond laser pulses with a \SI{2}{\kilo\hertz} repetition rate and \SI{300}{\milli\watt} average output power.
Afterwards, the resonator blank is soldered on a brass post and then turned to a sphere using a computer-controlled laser-lathe with the same fs-laser. For this contribution, we fabricated two different WGRs. The thick resonator has a major radius of $R=\SI{1.2}{\milli\meter}$, a minor radius of $\rho=\SI{380}{\micro\meter}$ and a thickness of $d_{z}=\SI{300}{\micro\meter}$, thus exhibiting free spectral ranges (FSRs) of $\Delta\nu_\mathrm{FSR}=18.1$~GHz for e-pol. and $\Delta\nu_\mathrm{FSR}=17.3$~GHz for o-pol. light at $\lambda_\mathrm{L}=1040$~nm.
The intrinsic \textit{Q}-factor is determined to be $Q = 2.4\times10^8$, which is limited  by material absorption only\,\cite{Leidinger15}.
To achieve larger tuning for a given voltage, we fabricated a second, thinner resonator with $d_{z}=\SI{70}{\micro\meter}$.
It possesses a major radius of $R=\SI{1}{\milli\meter}$ and a minor radius of $\rho=\SI{600}{\micro\meter}$, corresponding to FSRs of $\Delta\nu_\mathrm{FSR}= \SI{21.7}{\giga\hertz}$ (e-pol.) and $\Delta\nu_\mathrm{FSR}= \SI{20.8}{\giga\hertz}$ (o-pol.) at $\lambda_\mathrm{L}= \SI{1040}{\nano\meter}$. This resonator has an intrinsic quality factor of $Q=7\times10^7$. We suspect the absorption by the chromium electrode to be the main reason for the reduction in $Q$ compared with that of the thick resonator\cite{PhysRevA.94.053815}.
%%Experiment
Our setup is illustrated in Fig.\,\ref{fig:setup}.
To cover different wavelength regimes, where we expect different frequency shifts, we use the following continuous wave laser sources: a grating-stabilized diode laser emitting at around 1040$\,$nm, a Ti:sapphire laser at $800\,$nm wavelength or a DFB laser diode at $\SI{1.57}{\micro\meter}$ wavelength.
The laser light is then coupled into a fiber with a polarization controller.
This light leaving the fiber via a gradient index lens is focused into the prism.
Hence, an evanescent light field is generated on the base of the rutile prism. 
In order to change the coupling strength we can adjust the gap between the prism and the resonator with a piezo actuator.
To generate a sufficient signal level on the detector, but to avoid nonlinear-optical activity, we employ pump powers below $\SI{1}{\milli\watt}$ with up to $40\,\%$ coupling efficiency.
The top and bottom of the chromium-coated resonator are electrically connected to an arbitrary-function generator providing 20$\,$V maximum output voltage with less than $10\,$ns rise and fall time.
The shifted light is coupled out via the rutile prism and superimposed with the pump light that is not coupled in and reflected on the prism base. These two collinear waves are focused on a photodetector that is connected to a $12.5\text{-}\si{\giga\hertz}$-oscilloscope.
The whole setup is placed in a box made of acrylic glass providing a temperature stabilization of $\SI{1}{\milli\kelvin}$.

As illustrated in Fig.~\ref{fig:setup}, the fundamental resonator mode is not between the electrodes of the resonator.
Consequently, we cannot assume a simple plate capacitor model.
As we move outside the edges of the electrodes of the resonator, the electric field gets weaker.
For a smaller minor radius, the rim will protrude further outside the edges of the electrodes, reducing the effective electric field strength at the rim where the modes travel.
By simulating the electric field $E_z$ which points parallel to the z-axis (Fig. \ref{fig:setup}) and using our actual resonator geometries with COMSOL Multiphysics\textsuperscript{\textregistered}, we can determine the strength of the electric field at the position of the fundamental resonator mode, see  Fig.~\ref{fig:setup}~b), close up.
Since for higher values of $q$ and $p$ the mode profile changes significantly, our simple model might be less suited\cite{Breunig16}.
For the big resonator the field is reduced by a factor $\alpha_1 = 0.83$, and for the thinner resonator by $\alpha_2 = 0.97$. 
Thus, when we apply $\SI{1}{\volt}$ the electric field is assumed to be $E_\textrm{eff,1} = E_{z} \alpha_1 =\SI[per-mode=symbol]{2.8}[\,]{\volt\per\milli\meter}$ for the big and $E_\textrm{eff,2} = E_{z} \alpha_2 = \SI[per-mode=symbol]{13.9}[\,]{\volt\per\milli\meter}$ for the thin resonator.

\section{Results and discussion}

\begin{figure}[htb!]
	\centering
	\includegraphics{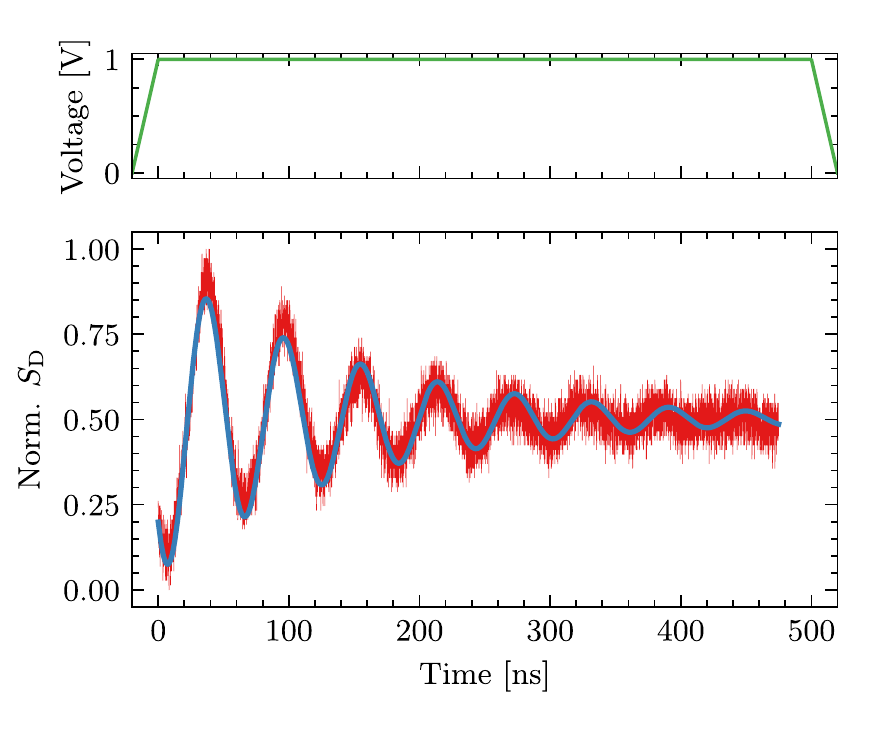}
	\caption{A square electrical signal~(green) of \SI{1}{\volt} amplitude applied to the electrodes of the resonator with $d_{z}=\SI{300}{\micro\meter}$ leads to a frequency shift of $\Delta\nu =\SI{17 \pm 0.1}{\mega\hertz}$ for o-pol. light. The decay time is determined to be $\tau = \SI{133}{\nano\second}$ (red: data, blue: fit of Eq.~(\ref{eq:theosignal})).}
	\label{fig:decay}
\end{figure}

First, we measure the beat note when we apply a periodic square signal with a frequency of $10\,$kHz and an on-time of $500\,$ns with an amplitude of $\SI{1}{\volt_{pp}}$ to the resonator.
The signal measured by the photodiode as well as a fit of Eq.~(\ref{eq:theosignal}) to the data is shown in Fig.~\ref{fig:decay}.
We determined a decay time of $\tau = \SI{133}{\nano\second} $.
This coincides according to Eq. (\ref{decaytime}) with the previously determined intrinsic quality factor.
We obtain a beat frequency of $\Delta\nu=\SI{17 \pm 0.1}{\mega\hertz}$.
The expected value for $\Delta\nu$ according to Eq.(\ref{eq:frequency_shift}) for the effective electric field $E_\textrm{eff,1} = \SI[per-mode=symbol]{2.8}{\volt\per\milli\meter}$ and $r_{13}\approx\SI[per-mode=symbol]{8.1}{\pico\meter\per\volt}$ is $\Delta\nu=\SI{16.1}{\mega\hertz}$.

\begin{figure}[ht]
	\centering
	\includegraphics{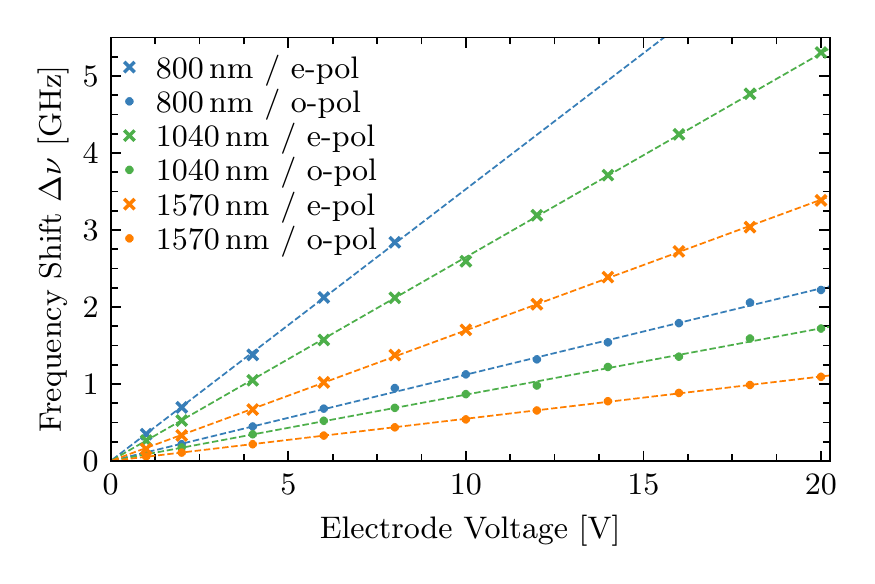}
	\caption{Experimental results for the obtained tuning for different light polarizations and for different applied voltages on the resonator with $d_{z}=\SI{70}{\micro\meter}$.
	The slope for e-pol.$\,$light is three times larger than that for o-pol.$\,$light due to different electro-optic coefficients $r_{\tiny\textrm{33}}\approx 3\times r_{\tiny\textrm{13}}$.}
	\label{fig:field_vs_tuning}
\end{figure}

\begin{figure}[hb]
\centering
	\begin{tikzpicture}
	\node[anchor=south west,inner sep=0] at (0,0) {\includegraphics{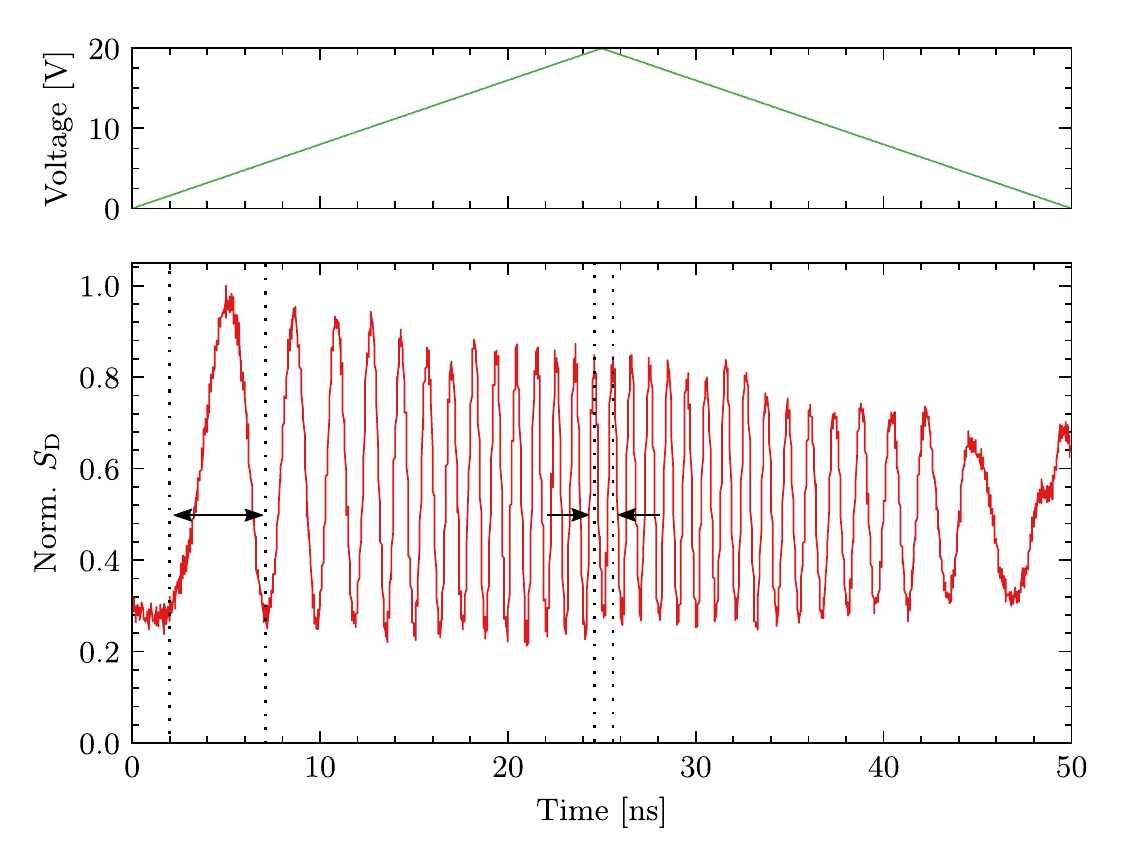}};
	\draw (2.3,6.4) node [] {$\Delta t_1$};
	\draw (6.15,6.4) node [] {$\Delta t_2$};
	\end{tikzpicture}
	\caption{Applying a triangular electrical signal\,(green) to the electrodes of the resonator with $d_{z}=\SI{300}{\micro\meter}$ leads to a positive frequency chirp and once the maximum voltage has been passed, a negative one, in this case shown for e-pol. light out-coupled from the resonator.
		We measured a period length of $\Delta t_1 \approx \SI{5}{\nano\second}$ and $\Delta t_2 \approx \SI{1}{\nano\second}$. This corresponds to beat frequencies $\Delta \nu_{\textrm{1}} = \SI{0.2}{\giga\hertz}$ and $\Delta \nu_{\textrm{2}} = \SI{1}{\giga\hertz}$, respectively.}
	\label{fig:chirp}
	\vspace{-0.5cm}
\end{figure}

In order to accomplish a larger tuning with the same voltage supply we use a thin resonator with $d_{z}=\SI{70}{\micro\meter}$.
We repeat the same measurement procedure as presented before for different voltages, wavelengths and polarizations.
The resulting frequency shifts are plotted in Fig.~\ref{fig:field_vs_tuning}.
Our experimental observation agrees with the expected linear behavior.
Using a linear fit we determine the frequency shift per Volt, which is for light of the wavelength $\lambda_\textrm{L} =\SI{1040}{\nano\meter}$ either $\SI[per-mode=symbol]{266 \pm 1}{\mega\hertz\per\volt}$~(e-pol.) or $\SI[per-mode=symbol]{86\pm 1}{\mega\hertz\per\volt}$~(o-pol.)
To compare this result with the theoretical value we calculate the frequency shift per Volt with Eq.~(\ref{eq:frequency_shift}) for the effective electric field again, by employing the $r$-coefficients for undoped CLN.
We obtain for e-pol. light $\SI[per-mode=symbol]{226.7}{\mega\hertz\per\volt}$ and for o-pol. light $\SI[per-mode=symbol]{80.5}{\mega\hertz\per\volt}$.
This measurement and the previous one suggest that the $r$-coefficient is higher than assumed.
%{\color{red}We determine them as $r_{13}\approx \SI[per-mode=symbol]{9}{\pico\meter\per\volt}$ and a $r_{33}\approx\SI[per-mode=symbol]{29}{\pico\meter\per\volt}$ at $\lambda_\mathrm{L}=\SI{1040}{\nano\meter}$.}
For the e-pol. laser light at the wavelength $\lambda_\textrm{L} = \SI{800}{\nano\meter}$ we observe a slope of $\SI[per-mode=symbol]{357\pm 2}{\mega\hertz\per\volt}$, which is almost twice the shift at $\lambda_\textrm{L}= \SI{1570}{\nano\meter}$, $\SI[per-mode=symbol]{168.8 \pm 0.3}{\mega\hertz\per\volt}$.
The same ratio holds for the orthogonal polarization. Here the two values are $\SI[per-mode=symbol]{112 \pm 1}{\mega\hertz\per\volt}$ and $\SI[per-mode=symbol]{54.9 \pm 0.1}{\mega\hertz\per\volt}$.
This also agrees with our expectations.
Neglecting small differences for the refractive index $n$ and the electro-optic coefficient $r$ at these two different wavelengths, we obtain twice the frequency shift by doubling of the laser frequency (Eq.~(\ref{eq:frequency_shift})).
So far, we used a square electrical signal form only. We studied also the influence of a triangular electrical signal applied to the electrodes of the $300\text{-}\si{\micro\meter}\text{-thick}$ resonator with a frequency of $\SI{10}{\kilo\hertz}$ and $\SI{50}{\nano\second}$ of on-time.
Since the electric field is changing its strength linearly with time, the frequency of the out-coupled light shifts linearly as well. The frequency chirp and the electric signal are shown in Fig.~\ref{fig:chirp}.
At the beginning and the end of the triangular electric signal, a beat frequency $\Delta\nu\approx\SI{0.2}{\giga\hertz}$ is observed.
The maximum frequency $\Delta\nu\approx\SI{1}{\giga\hertz}$ is reached at the maximum voltage of $\SI{20}{\volt}$.
After reaching this, we obtain a negative frequency chirp for the negative side of the triangular signal. 
\FloatBarrier
\section{Outlook and conclusion}

The presented method was realized with bulk lithium niobate whispering-gallery resonators. Despite this approach, a transfer to batch-processed chip-integrated resonators is feasible as we judge from recent advances of this technology platform \cite{Wolf17,Wang.2018}. Thinner resonators with lithium-niobate-on-insulator providing a thickness of only $\SI{2}{\micro\meter}$ allow tuning over tens of $\si{\giga\hertz}$ with considerably lower voltage.
Many applications can be envisaged. Just as one example, AFC is a possible candidate to serve for frequency modulated continuous wave LIDAR systems.

To conclude, we demonstrated adiabatic frequency conversion in an ultrahigh-$Q$ microresonator made out of lithium niobate by employing the Pockels effect.
We have realized AFC for different wavelengths and reach a maximum frequency shift of more than $\SI{5}{\giga\hertz}$ tuning for e-pol. light and of more than \SI{2}{\giga\hertz} for o-pol. light.
Our observed frequency shifts are considerably smaller compared to the $\SI{300}{\giga\hertz}$ reported before\cite{Preble.b}, but this tuning scheme affects the $Q$-factor of the resonator.
Tuning by the AC-Kerr effect\cite{Yoshiki.} does not affect the $Q$-factor of the resonator, but the achieved tuning is with about 150 MHz one order of magnitude smaller than the Pockels effect-based AFC as it is demonstrated, here. Furthermore, in our scheme the tuning range is so far limited only by the maximum output voltage of the function generator.
It has been shown that lithium niobate can withstand strong electric fields of up to $\SI[per-mode=symbol]{65}[\,]{\kilo\volt\per\milli\meter}$~\cite{Luennemann03}, corresponding to a refractive index change of $\Delta n = 4.8\times 10^{-3}$ for o-pol. light.
Such a refractive index change would lead to frequency shifts of a few $\si{\tera\hertz}$.
Moreover, positive and negative frequency chirps have been demonstrated.
\section*{Funding}
Gisela and Erwin Sick Fellowship
\section*{Acknowledgments}
The authors thank S.J. Herr for his valuable contributions in the early stage of the presented work and B. Aatz and D. Rutsch for technical support.
%\end{doublespace}
\section*{Disclosures}
The authors declare no conflicts of interest.

%\section*{References}
%%%%%%%%%%%%%%%%%%%%%%% References %%%%%%%%%%%%%%%%%%%%%%%%
%%%%%%%%%% If using BibTeX:
\bibliography{literatur1}

\begin{thebibliography}{10}
\newcommand{\enquote}[1]{``#1''}

\bibitem{Boyd.2008}
R.~W. Boyd, \emph{Nonlinear optics} ({Academic Press}, 2008), 3rd ed.

\bibitem{Herr14b}
T.~Herr, V.~Brasch, J.~D. Jost, C.~Y. Wang, N.~M. Kondratiev, M.~L. Gorodetsky,
  and T.~J. Kippenberg, \enquote{Temporal solitons in optical microresonators,}
  {\protect\JournalTitle{Nature Photonics}} \textbf{8}, 145--152 (2014).

\bibitem{Furst10}
J.~U. F{\"u}rst, D.~V. Strekalov, D.~Elser, A.~Aiello, U.~L. Andersen,
  C.~Marquardt, and G.~Leuchs, \enquote{Low-threshold optical parametric
  oscillations in a whispering-gallery-mode resonator,}
  {\protect\JournalTitle{Physical Review Letters}} \textbf{105}, 263904 (2010).

\bibitem{Breunig16}
I.~Breunig, \enquote{Three-wave mixing in whispering gallery resonators,}
  {\protect\JournalTitle{Laser {\&} Photonics Reviews}} \textbf{10}, 569--587
  (2016).

\bibitem{Notomi06}
M.~Notomi and S.~Mitsugi, \enquote{Wavelength conversion via dynamic refractive
  index tuning of a cavity,} {\protect\JournalTitle{Phys. Rev. A}} \textbf{73},
  051803 (2006).

\bibitem{Savchenkov07e}
A.~A. Savchenkov, A.~B. Matsko, V.~S. Ilchenko, and L.~Maleki, \enquote{Optical
  resonators with ten million finesse,} {\protect\JournalTitle{Optics Express}}
  \textbf{15}, 6768--6773 (2007).

\bibitem{Preble.2012}
S.~Preble, L.~Cao, A.~Elshaari, A.~Aboketaf, and D.~Adams, \enquote{Single
  photon adiabatic wavelength conversion,} {\protect\JournalTitle{Applied
  Physics Letters}} \textbf{101}, 171110 (2012).

\bibitem{Fan16b}
L.~Fan, C.-L. Zou, M.~Poot, R.~Cheng, X.~Guo, X.~Han, and H.~X. Tang,
  \enquote{Integrated optomechanical single-photon frequency shifter,}
  {\protect\JournalTitle{Nature Photonics}} \textbf{10}, 766--770 (2016).

\bibitem{Tanabe.}
T.~Tanabe, E.~Kuramochi, H.~Taniyama, and M.~Notomi, \enquote{Electro-optic
  adiabatic wavelength shifting and {$Q$} switching demonstrated using a p-i-n
  integrated photonic crystal nanocavity,} {\protect\JournalTitle{Opt. Lett.}}
  \textbf{35}, 3895--3897 (2010).

\bibitem{Tanabe.b}
T.~Tanabe, M.~Notomi, H.~Taniyama, and E.~Kuramochi, \enquote{Dynamic release
  of trapped light from an ultrahigh-{$Q$} nanocavity via adiabatic frequency
  tuning,} {\protect\JournalTitle{Physical Review Letters}} \textbf{102},
  043907 (2009).

\bibitem{Kondo18}
K.~Kondo and T.~Baba, \enquote{Adiabatic wavelength redshift by dynamic carrier
  depletion using $p\ensuremath{-}i\ensuremath{-}n$-diode--loaded photonic
  crystal waveguides,} {\protect\JournalTitle{Phys. Rev. A}} \textbf{97},
  033818 (2018).

\bibitem{Kampfrath10}
T.~Kampfrath, D.~M. Beggs, T.~P. White, A.~Melloni, T.~F. Krauss, and
  L.~Kuipers, \enquote{Ultrafast adiabatic manipulation of slow light in a
  photonic crystal,} {\protect\JournalTitle{Phys. Rev. A}} \textbf{81}, 043837
  (2010).

\bibitem{Upham.2010}
J.~Upham, Y.~Tanaka, T.~Asano, and S.~Noda, \enquote{On-the-fly wavelength
  conversion of photons by dynamic control of photonic waveguides,}
  {\protect\JournalTitle{Applied Physics Express}} \textbf{3}, 062001 (2010).

\bibitem{Kondo.14}
K.~Kondo and T.~Baba, \enquote{Dynamic wavelength conversion in copropagating
  slow-light pulses,} {\protect\JournalTitle{Physical Review Letters}}
  \textbf{112}, 223904 (2014).

\bibitem{Kabakova12}
I.~V. Kabakova, Z.~Yu, D.~Halliwell, P.-Y. Fonjallaz, O.~Tarasenko, C.~M.
  de~Sterke, and W.~Margulis, \enquote{Switching and dynamic wavelength
  conversion in a fiber grating cavity,} {\protect\JournalTitle{Journal of the
  Optical Society of America B}} \textbf{29}, 155 (2012).

\bibitem{Preble.b}
S.~F. Preble, Q.~Xu, and M.~Lipson, \enquote{Changing the colour of light in a
  silicon resonator,} {\protect\JournalTitle{Nature Photonics}} \textbf{1},
  293--296 (2007).

\bibitem{Yoshiki.}
W.~Yoshiki, Y.~Honda, M.~Kobayashi, T.~Tetsumoto, and T.~Tanabe,
  \enquote{Kerr-induced controllable adiabatic frequency conversion in an
  ultrahigh-{$Q$} silica toroid microcavity,} {\protect\JournalTitle{Optics
  Letters}} \textbf{41}, 5482--5485 (2016).

\bibitem{Faneaar4994}
L.~Fan, C.-L. Zou, R.~Cheng, X.~Guo, X.~Han, Z.~Gong, S.~Wang, and H.~X. Tang,
  \enquote{Superconducting cavity electro-optics: A platform for coherent
  photon conversion between superconducting and photonic circuits,}
  {\protect\JournalTitle{Science Advances}} \textbf{4}, eaar4994 (2018).

\bibitem{Rueda:16}
A.~Rueda, F.~Sedlmeir, M.~C. Collodo, U.~Vogl, B.~Stiller, G.~Schunk, D.~V.
  Strekalov, C.~Marquardt, J.~M. Fink, O.~Painter, G.~Leuchs, and H.~G.~L.
  Schwefel, \enquote{Efficient microwave to optical photon conversion: an
  electro-optical realization,} {\protect\JournalTitle{Optica}} \textbf{3},
  597--604 (2016).

\bibitem{dam2012room}
J.~S. Dam, P.~Tidemand-Lichtenberg, and C.~Pedersen, \enquote{Room-temperature
  mid-infrared single-photon spectral imaging,} {\protect\JournalTitle{Nature
  Photonics}} \textbf{6}, 788--793 (2012).

\bibitem{rueda2019resonant}
A.~Rueda, F.~Sedlmeir, M.~Kumari, G.~Leuchs, and H.~G. Schwefel,
  \enquote{Resonant electro-optic frequency comb,}
  {\protect\JournalTitle{Nature}} \textbf{568}, 378--381 (2019).

\bibitem{Zhang.2019}
M.~Zhang, B.~Buscaino, C.~Wang, A.~Shams-Ansari, C.~Reimer, R.~Zhu, J.~M. Kahn,
  and M.~Lon{\v{c}}ar, \enquote{Broadband electro-optic frequency comb
  generation in a lithium niobate microring resonator,}
  {\protect\JournalTitle{Nature}} \textbf{568}, 373--377 (2019).

\bibitem{dong2008inducing}
P.~Dong, S.~F. Preble, J.~T. Robinson, S.~Manipatruni, and M.~Lipson,
  \enquote{Inducing photonic transitions between discrete modes in a silicon
  optical microcavity,} {\protect\JournalTitle{Phys. Rev. Lett.}} \textbf{100},
  033904 (2008).

\bibitem{Daniel:11}
B.~A. Daniel, D.~N. Maywar, and G.~P. Agrawal, \enquote{Dynamic mode theory of
  optical resonators undergoing refractive index changes,}
  {\protect\JournalTitle{J. Opt. Soc. Am. B}} \textbf{28}, 2207--2215 (2011).

\bibitem{PhysRevA.94.053815}
C.~Javerzac-Galy, K.~Plekhanov, N.~R. Bernier, L.~D. Toth, A.~K. Feofanov, and
  T.~J. Kippenberg, \enquote{On-chip microwave-to-optical quantum coherent
  converter based on a superconducting resonator coupled to an electro-optic
  microresonator,} {\protect\JournalTitle{Phys. Rev. A}} \textbf{94}, 053815
  (2016).

\bibitem{Saleh2007}
B.~E.~A. Saleh and M.~C. Teich, \emph{Fundamentals of Photonics} (Wiley, 2007),
  chap. 20.1, p. 836, 2nd ed.

\bibitem{Mendez99b}
A.~M{\'e}ndez, A.~Garc{\'i}a-Caba{\~n}es, E.~Di{\'e}guez, and J.~M. Cabrera,
  \enquote{Wavelength dependence of electro-optic coefficients in congruent and
  quasi-stoichiometric {LiNbO}$_3$,} {\protect\JournalTitle{Electronics
  Letters}} \textbf{35}, 498--499 (1999).

\bibitem{Weis85}
R.~S. Weis and T.~K. Gaylord, \enquote{Lithium niobate: Summary of physical
  properties and crystal structure,} {\protect\JournalTitle{Appl. Phys. A}}
  \textbf{37}, 191--203 (1985).

\bibitem{Leidinger15}
M.~Leidinger, S.~Fieberg, N.~Waasem, F.~K{\"u}hnemann, K.~Buse, and I.~Breunig,
  \enquote{Comparative study on three highly sensitive absorption measurement
  techniques characterizing lithium niobate over its entire transparent
  spectral range,} {\protect\JournalTitle{Optics Express}} \textbf{23},
  21690--21705 (2015).

\bibitem{Wolf17}
R.~Wolf, I.~Breunig, H.~Zappe, and K.~Buse, \enquote{Cascaded second-order
  optical nonlinearities in on-chip micro rings,} {\protect\JournalTitle{Optics
  Express}} \textbf{25}, 29927--29933 (2017).

\bibitem{Wang.2018}
C.~Wang, M.~Zhang, X.~Chen, M.~Bertrand, A.~Shams-Ansari, S.~Chandrasekhar,
  P.~Winzer, and M.~Lon{\v{c}}ar, \enquote{Integrated lithium niobate
  electro-optic modulators operating at {CMOS}-compatible voltages,}
  {\protect\JournalTitle{Nature}} \textbf{562}, 101--104 (2018).

\bibitem{Luennemann03}
M.~Luennemann, U.~Hartwig, G.~Panotopoulos, and K.~Buse, \enquote{Electrooptic
  properties of lithium niobate crystals for extremely high external electric
  fields,} {\protect\JournalTitle{Applied Physics B}} \textbf{76}, 403--406
  (2003).

\end{thebibliography}

\end{document}